\documentclass [12pt]{iopart}
\usepackage{epsf}

\date{}
\begin{document}
\title{The effect of coordinate and momentum uncertainties on collision of 
coherent electrons}
\author{C.V. Usenko, V.O. Gnatovskyy and N. A. Cherkashina}

\address{National Shevchenko University of Kyiv, Ukraine}
\mailto{hnat@univ.kiev.ua}

\begin{abstract}

	We have examined the head-on collision of two electrons in approximation of 
	coherent states. We have shown that the character of collision depends 
	mainly on ratio of initial relative electron's momentum to momentum 
	uncertainty of electrons. When this ratio becomes greater then 1, the 
	Coulomb interaction does not practically influence the scattering.
\end{abstract}

\section{Introduction}

It is well known that solution of the electron-electron scattering problem, 
both classical and quantum, is generally reduced to finding the dependence 
of differential cross-section on the particles deflection angle. At that, 
the case of head-on collision is principally ignored despite the fact that 
it is in these scatterings quantum properties of electrons reveal themselves 
maximally. This is related first of all to their extreme mutual closing 
in, and therefore the manifestations of anti-symmetry of two-particle wave 
function become maximal. Moreover, the influence of Coulomb repulsion 
energy on two-particle system's behavior must be most significant. In this 
paper we have used the coherent states as one-particle states and have 
examined the head-on collision of coherent electrons in triplet state in 
center-of-mass system.

\section{Kinetic and interaction energy of two coherent electrons}

Let us assume that at an instant $t = 0$ the first electron is in coherent 
state $\left| \vec {\alpha } \right\rangle $ and the second one is in 
coherent state $\left| { - \vec {\alpha }} \right\rangle $. States $\left| 
\vec {\alpha } \right\rangle $ and $\left| { - \vec {\alpha }} \right\rangle 
$ are eigenstates for operator $\vec {a}$ with eigenvalues $\vec {\alpha }$ 
and $ - \vec {\alpha }$ correspondingly:

\[
\vec {a}\left| {\pm \vec {\alpha }} \right\rangle = \pm \vec {\alpha }\left| 
{\pm \vec {\alpha }} \right\rangle ,
\quad
\vec {a} = \frac{\sigma }{\hbar }\vec {r} + i\frac{\vec {p}}{2\sigma },
\quad
\vec {\alpha } = \frac{\sigma }{\hbar }\left\langle {\vec {r}_\alpha } 
\right\rangle + i\frac{\left\langle {\vec {p}_\alpha } \right\rangle 
}{2\sigma },
\]
\noindent
and may be written as \cite{we2}:

\[
\left| {\pm \vec {\alpha }} \right\rangle = \frac{1}{\left( {\sqrt {\sigma 
\sqrt {2\pi } } } \right)^3}\exp \left( { - \frac{\left( {\vec {p} \mp 
\left\langle {\vec {p}_\alpha } \right\rangle } \right)^2}{4\sigma ^2} \mp 
i\frac{\vec {p}\left\langle {\vec {r}_\alpha } \right\rangle }{\hbar }} 
\right),
\]
\noindent
where $\pm \left\langle {\vec {p}_\alpha } \right\rangle $ and $\pm 
\left\langle {\vec {r}_\alpha } \right\rangle $ are mean momentum and mean 
position of electrons at the instant $t = 0$, $\sigma $ is uncertainty of 
every component of particle's momentum in coherent states $\left| {\pm \vec 
{\alpha }} \right\rangle $:

\[
\sigma ^2 = \left\langle {p_i^2 } \right\rangle - \left\langle {p_i } 
\right\rangle ^2.
\]

Spatial part of two-particle state must be anti-symmetrical for triplet:

\begin{equation}
	\left| \Psi \right\rangle = \frac{\left| \vec {\alpha } \right\rangle \left| 
	{ - \vec {\alpha }} \right\rangle - \left| { - \vec {\alpha }} \right\rangle 
	\left| \vec {\alpha } \right\rangle }{\sqrt {2\left( {1 - \left| N 
	\right|^2} \right)} } \uparrow \uparrow ,
\label{state}
\end{equation}
\noindent
where $N = \left\langle {\vec {\alpha }} \mathrel{\left| {\vphantom {\vec 
{\alpha } { - \vec {\alpha }}}} \right. \kern-\nulldelimiterspace} { - \vec 
{\alpha }} \right\rangle $ is overlapping integral, $\left| N \right|^2 = 
\exp \left( { - 4\left| \vec {\alpha } \right|^2} \right)$. At a given 
instant $t$:

\[
\left| {\Psi ,t} \right\rangle = \exp \left( { - i\frac{H_0 }{\hbar }t} 
\right)\left| \Psi \right\rangle .
\]

This state satisfies Schroedinger equation:

\[
i\hbar \frac{d}{dt}\left| {\Psi ,t} \right\rangle = H_0 \left| {\Psi ,t} 
\right\rangle ,
\quad
H_0 = \frac{\vec {p}_1^2 }{2m} + \frac{\vec {p}_2^2 }{2m}.
\]

Average value of kinetic energy in the state $\left| {\Psi ,t} \right\rangle 
$ comprises three terms \cite{we1,we3}:

\begin{equation}
	\left\langle T \right\rangle = T_{cl} + T_{conf} + T_{corr} = 
	\frac{\left\langle {\vec {p}_\alpha } \right\rangle ^2}{m} + \frac{3\sigma 
	^2}{m} + \frac{4\sigma ^2\left| \vec {\alpha } \right|^2\left| N 
	\right|^2}{m\left( {1 - \left| N \right|^2} \right)}.
\end{equation}

The first term corresponds to free motion of classical particles, the second 
(energy of configuration) appears to be the contribution of momentum 
uncertainty; the third one (energy of correlation) appears due to 
overlapping of two non-orthogonal one-particle states.

Average value of Coulomb repulsion energy in the triplet state $\left| {\Psi 
,t} \right\rangle $

\[
\left\langle V \right\rangle = \left\langle {\Psi ,t} \right|\frac{e_0^2 
}{\left| {\vec {r}_1 - \vec {r}_2 } \right|}\left| {\Psi ,t} \right\rangle 
,
\]
\noindent
is equal \cite{we3} to:

\begin{equation}
	\left\langle V \right\rangle = e_0^2 \frac{\frac{1}{\left| \vec {R} 
	\right|}erf\left( {\frac{\sigma \left| \vec {R} \right|}{\hbar \sqrt {1 + 
	\omega ^2t^2} }} \right) + \frac{i}{\left| \vec {L} \right|}erf\left( 
	{\frac{i\sigma \left| \vec {L} \right|}{\hbar \sqrt {1 + \omega ^2t^2} }} 
	\right)\left| N \right|^2}{1 - \left| N \right|^2},
\end{equation}
\noindent
where

\[
\vec {R} = 2\left( {\left\langle {\vec {r}_\alpha } \right\rangle + 
\frac{\left\langle {\vec {p}_\alpha } \right\rangle }{m}t} \right),
\quad
\vec {L} = \frac{\hbar }{\sigma ^2}\left\langle {\vec {p}_\alpha } 
\right\rangle - 2\left\langle {\vec {r}_\alpha } \right\rangle \omega t,
\quad
\omega = \frac{2\sigma ^2}{\hbar m}.
\]

\section{Head-on collision of two coherent electrons}

Let us assume, that at the instant $t = 0$ the first electron is in 
coherent state $\left| \vec {\alpha } \right\rangle $ and has mean momentum 
$\vec {p}_1 \left( {t = 0} \right) = \left\langle {\vec {p}_\alpha } 
\right\rangle $, and mean position $\vec {r}_1 \left( {t = 0} \right) = 
\left\langle {\vec {r}_\alpha } \right\rangle $; the second electron is in 
coherent state $\left| { - \vec {\alpha }} \right\rangle $ with mean 
momentum and mean position $\vec {p}_2 \left( {t = 0} \right) = - 
\left\langle {\vec {p}_\alpha } \right\rangle $, $\vec {r}_2 \left( {t = 0} 
\right) = - \left\langle {\vec {r}_\alpha } \right\rangle $, respectively. 
As far as we examine the case of collision, the initial momentum is negative for 
the first electron and positive for the second.

As the solution of Schroedinger equation in zero-order approximation we 
have used the state (\ref{state}) in which $\left\langle {\vec {p}_\alpha } 
\right\rangle $ and $\left\langle {\vec {r}_\alpha } \right\rangle $ depend 
upon time according to Hamilton equations: 

\begin{equation}
	\frac{dr}{dt} = \frac{1}{2}\frac{\partial H}{\partial p},
	\quad
	\frac{dp}{dt} = - \frac{1}{2}\frac{\partial H}{\partial r},
\label{H}
\end{equation}
\noindent
where 

\[
H = \left\langle T \right\rangle + \left\langle V \right\rangle ,
\quad
p = \left\langle {\vec {p}_\alpha } \right\rangle ,
\quad
r = \left\langle {\vec {r}_\alpha } \right\rangle .
\]

In \cite{we3} it is shown that $\left\langle V \right\rangle $ can be considered 
as perturbation in wide enough range of $\left\langle {\vec {p}_\alpha } 
\right\rangle $ and $\left\langle {\vec {r}_\alpha } \right\rangle $. Since 
analytical solution of the system of equations (\ref{H}) is not found we use its 
numerical solution. 

\begin{figure}[h]
\epsfbox{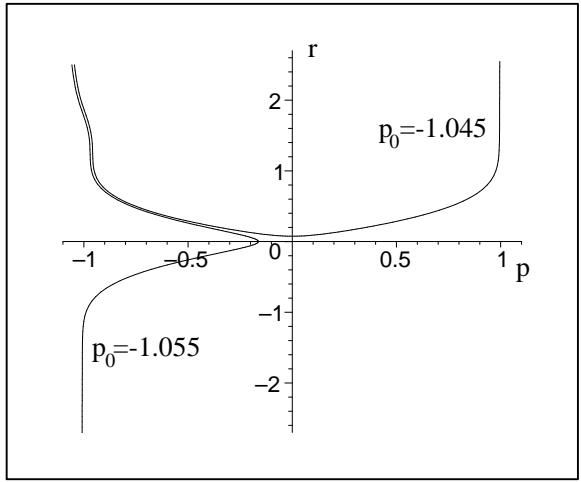}	
\caption{Phase trajectories for head-on collision. Initial distance between electrons is 5 a.u., momentum uncertainty of particles is $\sigma=1$ a.u.}
\label{f1}
\end{figure}

In Fig.\ref{f1} the two typical phase trajectories are shown for head-on collision 
of two coherent electrons. The distance between electrons and their momenta 
are in atomic units, initial distance is 5 a.u. One can see that according 
to the ratio of initial relative momentum to momentum uncertainty of each 
electron, two principally different kinds of behavior of phase trajectories 
are possible. As long as the initial momentum of electrons is less than some 
critical momentum $p_{cr} $ the effective deflection occurs. When initial 
momentum of electrons exceeds $p_{cr} $ the effective penetration is the 
case. Let us note that these two cases are experimentally indistinguishable due 
to identity of electrons.

\begin{figure}[h]
\epsfbox{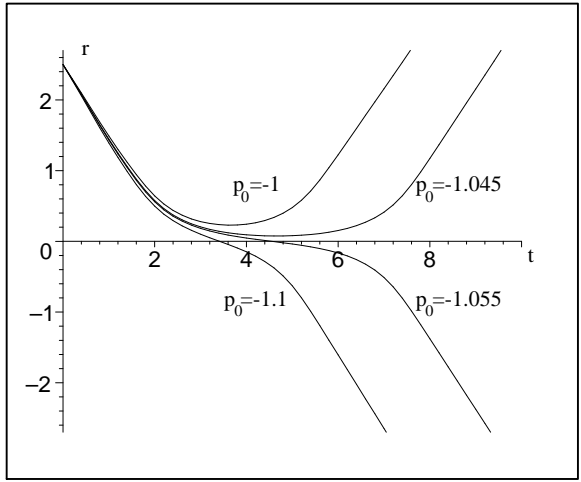}	
\caption{Dependence of first electron's posinion on time. Initial coordinate of the first electron is equal to 2.5 a.u., $\sigma=1$ a.u.}
\label{f2}
\end{figure}

Concerning the identity of particles the electron's \textit{return time} $t$ may be used as an 
observable feature of collision process. This term implies time span 
necessary to revert the distance between electrons to initial value. For calculation of $t$ 
the plots of dependence of electron's coordinate upon time are 
used. The example of such dependence is shown in Fig.\ref{f2} at the initial distance 
between electrons 5 a.u. It should be noticed that a significant delay 
near the origin is possible both in the case of deflection and penetration.

\begin{figure}[h]
\epsfbox{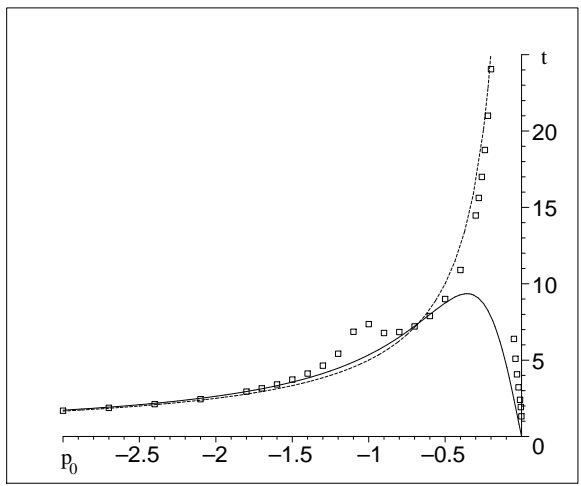}	
\caption{Dependence of return time of first electron on initial momentum $p_{0}$. Momentum uncertainty of electrons is $\sigma=1$ a.u. Initial distance between particles is 5 a.u.}
\label{f3}
\end{figure}

The dependence of \textit{return time} $t$ on initial momentum at the initial distance between particles 
 5 a.u. is shown in Fig.\ref{f3} with boxes. For comparison, analogous 
dependences for non-interacting particles (dashed line) and classical 
electrons (solid line) are given, and they almost coincide for large 
relative momenta. Fig.\ref{f3} shows that dependence of \textit{return time }$t$ on the initial momentum 
of electrons $p_0 $ coincides with that for classical electrons when 
${p_0 } \mathord{\left/ {\vphantom {{p_0 } {\sigma < 0.1}}} \right. 
\kern-\nulldelimiterspace} {\sigma < 0.1}$ only. In the range $0.2 < {p_0 } 
\mathord{\left/ {\vphantom {{p_0 } {\sigma < 1}}} \right. 
\kern-\nulldelimiterspace} {\sigma < 1}$ the Coulomb repulsion may be 
neglected, though as Fig.\ref{f1} shows, this range corresponds to effective 
deflection. In the area ${p_0 } \mathord{\left/ {\vphantom {{p_0 } {\sigma 
\approx 1}}} \right. \kern-\nulldelimiterspace} {\sigma \approx 1}$ a delay 
occurs. When initial momentum $p_0 $ is large, ${p_0 } \mathord{\left/ 
{\vphantom {{p_0 } \sigma }} \right. \kern-\nulldelimiterspace} \sigma > 1$, 
interaction may be ignored and phase trajectories accord with effective 
penetration.

\section{Conclusions}

The analysis of head-on collision of two coherent electrons has shown that 
the Coulomb repulsion is considerable in the area of small values of 
relative momenta only. For almost all values of initial relative momenta the 
system's behavior practically does not differ from classical. Essentially 
quantum scattering can be observed only in the area ${p_0 } \mathord{\left/ 
{\vphantom {{p_0 } {\sigma \approx 1}}} \right. \kern-\nulldelimiterspace} 
{\sigma \approx 1}$, when the distance between particles at the instant of 
stop is less than position uncertainty of electrons.

%\begin{thebibliography}{9}

%\end{thebibliography}

\end{document}